\documentclass[twocolumn,aps,pra,longbibliography,superscriptaddress,nofootinbib,10pt]{revtex4-1}

\usepackage{amsmath}
\usepackage{amsfonts}
\usepackage{amssymb}
\usepackage{graphicx}
\usepackage[dvipsnames]{xcolor}
\usepackage[colorlinks=true,citecolor=blue,linkcolor=blue,urlcolor=blue]{hyperref}
\usepackage{amstext}
\usepackage{latexsym}
\usepackage[running,mathlines]{lineno}

\providecommand{\moy}[1]{\langle #1 \rangle}
\providecommand{\bra}[1]{\langle #1 \rvert}
\providecommand{\ket}[1]{\lvert #1 \rangle}
\providecommand{\braket}[2]{\langle #1 \rvert #2 \rangle}

\providecommand{\be}{\begin{equation}}
\providecommand{\ee}{\end{equation}}
\providecommand{\ba}{\begin{eqnarray}}
\providecommand{\ea}{\end{eqnarray}}

\newcommand{\sz}{\sigma_{z}}
\newcommand{\sma}{\sigma_{+}}
\newcommand{\sme}{\sigma_{-}}

\begin{document}
\raggedbottom

\title{Quantum coherence and speed limit in the mean-field Dicke model of superradiance}
\author{D. Z. Rossatto}
\thanks{These authors contributed equally to this work.}
\affiliation{Universidade Estadual Paulista (Unesp), Campus Experimental de Itapeva, 18409-010 Itapeva, S\~{a}o Paulo, Brazil}
\author{D. P. Pires}
\thanks{These authors contributed equally to this work.}
\affiliation{Departamento de F\'{i}sica Te\'{o}rica e Experimental, Universidade Federal do Rio Grande do Norte, 59072-970 Natal, Rio Grande do Norte, Brazil}
\author{F. M. de Paula}
\affiliation{Centro de Ci\^{e}ncias Naturais e Humanas, Universidade Federal do ABC, 09606-070 S\~{a}o Bernardo do Campo, S\~{a}o Paulo, Brazil}
\author{O. P. de S\'{a} Neto}
\affiliation{Coordena\c{c}\~{a}o de Ci\^{e}ncia da Computa\c{c}\~{a}o, Universidade Estadual do Piau\'{i}, 64202-220 Parna\'{i}ba, Piau\'{i}, Brazil}
\affiliation{PPGQ, Universidade Estadual do Piau\'{i}, Rua Jo\~{a}o Cabral 2231, 64002-150 Teresina, Piau\'{i}, Brazil}

\begin{abstract}
Dicke superrandiance is a cooperative phenomenon which arises from the collective coupling of an ensemble of atoms to the electromagnetic radiation. Here we discuss the quantifying of quantum coherence for the Dicke model of superradiance in the mean-field approximation. We found the single-atom $l_1$-norm of coherence is given by the square root of the normalized average intensity of radiation emitted by the superradiant system. This validates quantum coherence as a useful figure of merit towards the understanding of superradiance phenomenon in the mean-field approach. In particular, this result suggests probing the single-atom coherence through the radiation intensity in superradiant systems, which might be useful in experimental realizations where is unfeasible to address atoms individually. Furthermore, given the nonlinear unitary dynamics of the time-dependent single-atom state that effectively describes the system of $N$ atoms, we analyze the quantum speed limit time and its interplay with the $l_1$-norm of coherence. We verify the quantum coherence speeds up the evolution of the superradiant system, i.e., the more coherence stored on the single-atom state, the faster the evolution. These findings unveil the role played by quantum coherence in superradiant systems, which in turn could be of interest for communities of both condensed matter physics and quantum optics.

\end{abstract}

%\pacs{03.65.Aa, 03.67.Mn, 03.65.Yz, 03.65.Ta}

\maketitle

%%%%%%%%%%%%%%%%%%%%%%%%%%%%%%%%%%%%%%%%%%%%%%%
%%%%%%%%%%%%%%---INTRODUCTION---%%%%%%%%%%%%%%%%%%%%%%
%%%%%%%%%%%%%%%%%%%%%%%%%%%%%%%%%%%%%%%%%%%%%%%

\section{Introduction}
\label{section0001}

Light-matter interaction plays a striking role in the understanding of several physical phenomena~\cite{PhysRevA.8.2517}, thus being a subject of significant interest to research on laser cooling and atomic trapping~\cite{RevModPhys.58.699,PhysRevLett.61.826,Chu:Wieman:89,RevModPhys.70.721}, cavity quantum electrodynamics~\cite{RevModPhys.73.565,RevModPhys.75.281}, and more recently quantum computing~\cite{PhysRevLett.92.127902}. It is noteworthy that the Dicke model, which des\-cri\-bes the coupling of a single mode of the radiation field with an ensemble of two-level systems, stands as a paradigmatic toy model from quantum optics~\cite{Dicke1954,PhysRevA.7.831,Garraway2011,Kirton2018}. In turn, the collective and coherent interaction can promote the well-known Dicke superradiance, in which the system spontaneously emits radiation at high intensity in a short time window~\cite{Nature_285_70_1980,Gross1982}. Remarkably, experimental rea\-li\-za\-tion of superradiance has been performed in a variety of quantum platforms~\cite{PhysRevLett.30.309,PhysRevLett.36.1035,PhysRevLett.76.2049,Inouye1999,NaturePhys_3_2,Rohlsberger2010,Mlynek2014,PhysRevLett.115.063601,PhysRevLett.114.023601,PhysRevLett.114.023602,PhysRevLett.117.210503,PhysRevLett.124.013601,Kim2017}. Moreover, exploiting the physical richness of the superradiant phase~\cite{Hepp1973,Nature_464_7293,PhysRevA.96.033633,PhysRevLett.125.050402}, superradiance has applications in ultra-narrow-linewidth
lasers~\cite{PhysRevLett.71.995,Bohnet2012}, quantum communication~\cite{Duan2001,Kimble2008}, sensitive gravimeters~\cite{Liao2015}, and quantum batteries~\cite{PhysRevLett.120.117702,PhysRevLett.122.047702,PhysRevB.99.205437}.

Motivated by the ubiquitous role of radiation intensity in the superradiance phenomenon, one can ask if Dicke superradiance somehow accelerates the evolution of the quantum system, the latter remaining as a challenge for the design of faster quantum information-processing devices~\cite{PhysRevLett.103.240501,LARS_2018}. In particular, this quantum signature could be captured by the so-called quantum speed limit (QSL), i.e., the minimum time of evolution required for a quantum system evolve between two given states~\cite{1945_JPhysURSS_9_249,1992_PhysicaD_120_188,2009_PhysRevLett_103_160502}. Nowadays, QSL find applications in quantum computation and quantum communication~\cite{PhysRevA.82.022318}, quantum metrology~\cite{PhysRevA.97.022109}, and quantum thermodynamics~\cite{PhysRevLett.118.150601}. Furthermore, Dicke superradiance has been recently addressed under the viewpoint of local quantum uncertainty quantifiers~\cite{PhysRevA.94.023819}, as well as quantum correlations~\cite{PhysRevLett.112.140402,PhysRevA.101.052310}. This motivates an investigation of the superradiance employing another figure of merit as quantum coherence, particularly focusing on its resource-theoretical approach~\cite{PhysRevLett.113.140401,RevModPhys.89.041003}. Quantum co\-he\-ren\-ce is a remar\-ka\-ble fingerprint of non-classical systems linked to the quantum superposition principle, which plays an essential role in quantum optics~\cite{PhysRev.130.2529}, quantum thermodyna\-mics~\cite{PhysRevA.93.052335}, condensed matter physics~\cite{PhysRevB.93.184428}, and bio\-lo\-gi\-cal systems~\cite{2013_829687}. In fact, it has been shown that coherence and superradiance are mutually interconvertible resources~\cite{PhysRevA.97.052304}. Moreover, one can investigate the interplay between the many-body coherences of Dicke superradiance and the QSL in multiparticle systems~\cite{PhysRevResearch.2.023125}.

In this paper, we discuss the quantifying of quantum coherence and quantum speed limit time, and also their interplay, for the Dicke model of superradiance in the mean-field approximation. We find quantum coherence is suddenly suppressed for a large number of atoms, while exhibits a maximum value at the time delay of superradiance (time of maximum intensity), at which in turn is robust to the number of atom increasing. Importantly, we show that $\ell_1$-norm of coherence is related to the intensity of radiation emitted by the superradiant system. This result suggests that probing the single-atom coherence through the radiation intensity in superradiant systems might be useful in experimental setups where is unfeasible to address atoms individually. It is noteworthy that the QSL bound saturates as one increases the number of atoms $N$ in the system, thus suggesting that quantum information-processing devices based on Dicke superradiance could operate at maximum speed in the limit $N \gg 1$. Moreover, we explore the relation of quantum coherence and QSL time, thus analyzing the former as a resource capable to speed up the unitary dynamics of each single two-level atom. 

The paper is organized as follows. In Sec.~\ref{sec:model} we briefly review the basic features of the Dicke model of superradiance in view of mean-field approximation. In Sec.~\ref{sec:quantumcoherence001} we discuss the role of quantum coherence in the referred model. In Sec.~\ref{sec:QSLsection0001} we study the quantum speed limit time with regard to the effective nonlinear unitary evolution of two non-orthogonal single-atom pure states. In addition, we investigate the interplay between the QSL bound and quantum coherence. Finally, we sumarize our main results and present the conclusions.

%%%%%%%%%%%%%%%%%%%%%%%%%%%%%%%%%%%%%%%%%%%%%%%
%%%%%%%%%%%%%%%%%%%---SECTION II---%%%%%%%%%%%%%%%%%%%%
%%%%%%%%%%%%%%%%%%%%%%%%%%%%%%%%%%%%%%%%%%%%%%%

\section{Physical System}
\label{sec:model}

Let us consider the Dicke model of superradiance, a system of $N$ identical two-level atoms with transition frequency $\omega$, which interacts collectively with their surrounding electromagnetic field in the vacuum state (zero temperature)~\cite{Benedict1996,Garraway2011}. For transitions between Dicke states~\cite{MandelWolf}, considering $N \gg 1$ and the system weakly coupled to its environment, the dynamics of this system can be described by the Lindblad-type mean-field master equation ($\hbar=1$)~\cite{Gross1982,BreuerPet}
\begin{equation}
\label{mecoll}
\frac{d{\rho_N}}{dt} = -i\omega[{J^z},{\rho_N}] - \frac{\gamma_0}{2}\left(\{ {J^+}{J^-},{\rho_N}\} - 2\, {J^-}{\rho_N}{J^+} \right) ~,
\end{equation}
where ${J^z} = (1/2)\, {\sum_{j=1}^N} {\sigma_j^z}$ and ${J^{\pm}} = {\sum_{j=1}^N} {\sigma^{\pm}_j}$ are collective operators, with ${\sigma_j^z}$ and ${\sigma^{\pm}_j}$ denoting the Pauli matrices associated with the $j$-th atom, $[X,Y] =XY - YX$ is the commutator, and $\{ X,Y \} = XY + YX$ is the anticommutator. In particular, this mean-field master equation can be mapped onto a nonlinear Schr\"{o}dinger-type equation by embedding the dy\-na\-mics of the $N$-atom mean-field state ${\rho_N} \approx {(|{\psi_t}\rangle\langle{\psi_t}|)^{\otimes N}}$ into an effective unitary dynamics of each two-level atom, described by $\ket{\psi_t}$, which in turn evolves accor\-ding to the nonlinear Hamiltonian $H_t = (\omega/2)\,{\sigma_z} + i \, ({N\gamma_0}/{2}) (\bra{\psi_t}\sma\ket{\psi_t}\sme - \bra{\psi_t}\sme\ket{\psi_t}\sma)$, where ${\sigma_+}$ (${\sigma_-}$) is the raising (lowering) operator, and $\gamma_0$ is the spontaneous emission rate of each single two-level atom~\cite{BreuerPet}. 

The solution for the nonlinear Schr\"{o}dinger equation $({d}/{dt})\ket{\psi_t} = -i {H_t}\ket{\psi_t}$ is described by the time-dependent single-atom state~\cite{BreuerPet}
\begin{equation}
\label{hamilteffective0002}
\ket{\psi_t} = \sqrt{1 - {p_t}} \, {e^{i\frac{\omega}{2} t}} \ket{g} + \sqrt{{p_t}} \, {e^{-i\frac{\omega}{2} t}} \ket{e} ~,
\end{equation}
where $|g\rangle$ ($|e\rangle$) stands for the ground (excited) state of the two-level atom. Here 
\begin{equation}
\label{hamilteffective0002001}
{p_t} = [{e^{\gamma_0 N (t - t_D)}+1}]^{-1}
\end{equation}
denotes the probability of finding a single atom in the excited state at time $t$, with $0 \leq {p_t} \leq 1$, while ${t_D} = {(\gamma_0} N)^{-1} \ln (N)$ stands for the time delay of the superradiance (time of maximum intensity)~\cite{Gross1982,Benedict1996}. Finally, with $\ket{\psi_t}$ given by Eq.~\eqref{hamilteffective0002}, the nonlinear Hamiltonian can be written as
\begin{equation}
\label{hamilteffective0001}
{H_t} = \frac{\omega}{2}\sz -i \frac{N\gamma_0}{2}\sqrt{{p_t}(1 - {p_t})}(\sma e^{-i\omega t}- \sme e^{i\omega t}) ~.
\end{equation}
% the density matrix solving the refereed master equation is written as a product of uncorrelated marginal pure states.
It is noteworthy that, given the probability distribution $p_t$ in Eq.~\eqref{hamilteffective0002001}, the average intensity of radiation emitted by the system of $N$ atoms can be written as $I(t) = -N\omega (dp_t/dt) = ({N^2}\omega{\gamma_0}/4) \, {\text{sech}^2}[(N{\gamma_0}/2)(t - {t_D}) ]$, which is proportional to $N^2$, characterizing the superradiance~\cite{BreuerPet}. For $t = {t_D}$, the system emits radiation with the maximum intensity value ${I_\text{max}} = I( t = {t_D}) = {N^2}\omega{\gamma_0}/4$.

In general, Eq.~(1) describes a system of two-level atoms in free space in the mean-field approximation, which in turn is valid in the limit $N \to \infty$~\cite{BreuerPet}. Nevertheless, it is worth mention that Eq.~(1) also describes an atomic cloud with a finite number $N$ of noninteracting two-level atoms coupled to a leaking cavity~\cite{Rossatto2011,PhysRevA.4.302,Delanty2011,Mlynek2014,PhysRevLett.124.013601} in the so-called bad-cavity limit~\cite{haroche2006exploring}. Just to clarify, in this case the cavity dissipation surpasses the effective coupling between the cavity mode and the atomic cloud. Therefore, our results might embody the case of a finite number of atoms.

%%%%%%%%%%%%%%%%%%%%%%%%%%%%%%%%%%%%%%%%%%%%%%%
%%%%%%%%%%%%%%%%---SECTION III---%%%%%%%%%%%%%%%%%%%%%%
%%%%%%%%%%%%%%%%%%%%%%%%%%%%%%%%%%%%%%%%%%%%%%%

\section{Quantum Coherence}
\label{sec:quantumcoherence001}

In this section we briefly introduce the main concepts of quantum coherence and discuss its role in the superradiance phenomenon addressed in Sec.~\ref{sec:model}. Over half decade ago, the seminal work by Baumgratz {\it et al.}~\cite{PhysRevLett.113.140401} introduced the minimal theoretical framework for the quantification of quantum coherence. This approach opened an avenue for the characterization of quantum coherence under the scope of resource theories, which currently still is a matter of intense debate~\cite{PhysRevA.94.052336,PhysRevA.95.019902,PhysRevLett.117.030401,PhysRevLett.116.120404,PhysRevX.6.041028}. 

Here we will briefly review the key aspects of quantifying quantum coherence discussed in Ref.~\cite{PhysRevLett.113.140401}. We shall consider a physical system defined on a $d$-dimensional Hilbert space $\mathcal{H}$ endowed with some re\-fe\-ren\-ce basis ${\{ | {j} \rangle \}_{j = 0}^{d - 1}}$. In turn, the state of the system is described by a density matrix $\rho \in \mathcal{D}(\mathcal{H})$, where $\mathcal{D}(\mathcal{H}) = \{ {\rho^{\dagger}} = \rho,~\rho\geq 0,~\text{Tr}(\rho) = 1\}$ stands for the convex space of po\-si\-ti\-ve semi-definite density operators. In particular, the subset $\mathcal{I}\in \mathcal{D}(\mathcal{H})$ of incoherent states encompass the family of density matrices as $\delta = {\sum_j}\, {q_j}|{j}\rangle\langle{j}|$ that are diagonal in the reference basis, with $0 \leq {q_j} \leq 1$ and ${\sum_j}\, {q_j} = 1$. In summary, a {\it bona fide} quantum cohe\-ren\-ce quantifier $C(\rho)$ must satisfy the following properties~\cite{PhysRevLett.113.140401,RevModPhys.89.041003} (i) non-negativity, i.e., $C(\rho) \geq 0$ for all state $\rho$, with $C(\rho) = 0$ iff $\rho \in \mathcal{I}$; (ii) con\-ve\-xi\-ty under mixing, i.e., $C({\sum_n}\, {q_n} {\rho_n}) \leq {\sum_n}\, {q_n} C({\rho_n})$, with ${\rho_n} \in \mathcal{D}(\mathcal{H})$, $0 \leq {q_n} \leq 1$, and ${\sum_n}\, {q_n} =1$; (iii) mo\-no\-to\-ni\-city under in\-co\-he\-rent completely positive and trace-preserving (ICPTP) maps, i.e., $C(\mathcal{E}[\rho]) \leq C(\rho)$, for all ICPTP map $\mathcal{E}[\bullet]$; (iv) strong monotonicity, i.e., $C(\rho) \geq {\sum_n} {q_n} C({\rho_n})$, where ${\rho_n} = {q_n^{-1}}{K_n} \rho {K^{\dagger}_n}$ sets the post-measured states for arbitrary Kraus o\-pe\-rators $\{ {K_n} \}$ satisfying ${\sum_n} {K^{\dagger}_n} {K_n} = \mathbb{I}$ and $K_n\mathcal{I}K^{\dagger}_{n}\subset \mathcal{I}$, with ${q_n} = \mathrm{Tr}({K_n} \rho {K^{\dagger}_n})$.

Some widely known quantum coherence measures include relative entropy of coherence~\cite{PhysRevLett.113.140401}, geometric co\-he\-ren\-ce~\cite{PhysRevLett.115.020403}, and robustness of coherence~\cite{PhysRevLett.116.150502}. In addition, the $l_1$-norm of coherence, which is defined in terms of $l_1$-distance between $\rho$ and its closest incoherent state, is of fundamental interest since its exact calculation is readily given by ${C}(\rho) = {\sum_{j,l (j\neq l)}}{| {\rho_{jl}} |}$, where $\rho_{jl}$ sets the off-diagonal ele\-ments of $\rho$ eva\-lua\-ted with respect to the re\-fe\-ren\-ce basis ${\{ | {j} \rangle \}_{j = 0}^{d - 1}}$~\cite{PhysRevLett.113.140401}. Importantly, for a pure single-qubit state (i.e., $d = 2$ and $\rho = |\phi\rangle\langle{\phi}|$), the referred quantum coherence measures are monotonically related to each other, and thus they shall exhibit the same qua\-li\-ta\-ti\-ve behavior~\cite{RevModPhys.89.041003}. In this context, without any loss of generality, from now on we will address the $l_1$-norm of co\-he\-ren\-ce to characterize the role of quantum coherence in the superradiant system previously discussed.

From Sec.~\ref{sec:model}, fixing the reference basis $\{ |{e}\rangle, |{g}\rangle \}$, Eq.~\eqref{hamilteffective0002} implies the single-atom density operator 
\begin{align}
{\rho_t} &= (1 - {p_t})|{g}\rangle\langle{g}| + {p_t}|{e}\rangle\langle{e}| \nonumber\\
&+ \sqrt{{p_t}(1 - {p_t})}({e^{i\omega t}}|{g}\rangle\langle{e}| + {e^{- i\omega t}}|{e}\rangle\langle{g}|) ~, 
\end{align}
with $p_t$ given in Eq.~\eqref{hamilteffective0002001}. In this case, $l_1$-norm of co\-he\-ren\-ce is given by
\begin{equation}
\label{eq:coherence00001}
{C}({\rho_t}) = \text{sech}\left( \frac{N{\gamma_0}}{2}\, (t - {t_D}) \right) ~.
\end{equation}
Note that ${C}({\rho_t})$ is a bell-shaped symmetric function over time $t$, centered at $t = t_D$, with a full width at half maximum scaling as $(\gamma_0 N)^{-1}$. For $t = 0$, Eq.~\eqref{eq:coherence00001} reduces to ${C}({\rho_0}) = \text{sech}\left( N{\gamma_0}{t_D}/{2}\right)$, which in turn approaches to zero in the limit $N \rightarrow \infty$ for ${t_D} \neq 0$.

%%%%%%%%%
%\begin{figure}[t]
%	\centering
%	\includegraphics[trim = 0mm 0mm 3mm 7mm, clip, width=0.75\linewidth]{Fig.pdf}
%	\caption{Density plot of $l_1$-norm of coherence ${C_{l_1}}({\rho_t})$ [Eq.~\eqref{eq:coherence00001}] as a function of the dimensionless parameters $\omega(t - t_D)$ and the parameter $N{\gamma_0}/(2\omega)$.}
%	\label{figure01}
%\end{figure}
%%%%%%%%%%

Figure~\ref{figure}(a) shows the density plot of ${C}({\rho_t})$ as a function of dimensionless parameters $\omega(t - {t_D})$ and $N{\gamma_0}/(2\omega)$. For $0 < t \le {t_D}$, ${C}({\rho_t})$ starts increasing monotonically and reaches its maximum value ${C}({\rho_{t_D}}) = 1$ at the time delay $t = {t_D}$. In fact, ${C}({\rho_{t_D}})$ is always maximum regardless the value of $N{\gamma_0}/(2\omega)$. For $t > {t_D}$, ${C}({\rho_t})$ decreases mono\-to\-nically and asymptotically approaches zero in time. In the very underdamped regime, i.e., $N{\gamma_0}/(2\omega) \ll 1$, the quantum co\-he\-rence remains appro\-xi\-mately constant around its maximum value ${C}({\rho_t}) \sim 1$, while for the very overdamped regime, $N{\gamma_0}/(2\omega) \gg 1$, it follows that ${C}({\rho_t}) \sim 0$ for all $t \neq {t_D}$. In other words, the more atoms the system has, the less quantum coherence is stored on each single-atom state and therefore on the $N$-atom state [see Eq.~\eqref{eq:coherence00001002}], except at time $t = t_D$. Roughly speaking, for $t = {t_D}$ the system of two-level atoms undergoes a constructive quantum interference yielding the cooperative effect of superradiance. In fact, the system will radiate with maximum intensity, in a short time window. However, for all $t \neq {t_D}$, the single-particle coherences of the atomic ensemble will be suppressed as the system becomes larger, and thus the quantum properties tend to be negligible for $N$ sufficiently large.

Quite interestingly, Eq.~\eqref{eq:coherence00001} can be written in terms of the average intensity of radiation, $I(t) = ({N^2}\omega{\gamma_0}/4) \, {\text{sech}^2}[(N{\gamma_0}/2)(t - {t_D})]$. Indeed, one readily concludes
\begin{equation}
\label{eq:coherence00002}
{C}({\rho_t}) = \sqrt{\frac{{I(t)} }{I_\text{max}}} ~,
\end{equation}
with $I_\text{max} = {N^2}\omega{\gamma_0}/4$ the maximum intensity value. From Eq.~\eqref{eq:coherence00002}, note the single-atom co\-he\-ren\-ce depends on the square root of the normalized average intensity coo\-pe\-ratively emitted by the whole system. 

We shall stress the intensity $I(t)$ immediately vanishes for the case in which quantum coherence $C(t)$ is identically zero. Therefore, quantum coherence, instead of quantum entanglement, stands as a figure of merit towards the understanding of the superradiance phenomenon in the mean-field approach. More fundamentally, the relation $C({\rho_t}) \propto \sqrt{I(t)}$ between coherence and radiation intensity becomes clearer when one rea\-li\-zes that for two-level atoms the $l_1$-norm of co\-he\-ren\-ce is given by the normalized microscopic dipole moment, i.e., $\moy{\sigma_x}$, which in turn stand as a natural measure of coherence in the process of collective spontaneous emission~\cite{PhysRevA.97.052304}. Indeed, remembering the average intensity of radiation can be written in terms of the coherence of the normalized total electric dipole moment, i.e., $I(t) \propto \moy{J^{+}J^{-}}$~\cite{Gross1982,Benedict1996}, one may readily verify that ${[C({\rho_t})]^2} \propto \moy{J^{+}J^{-}}$, thus validating the $l_1$-norm of coherence as a figure of merit of the superradiance phenomenon in the mean-field approach.

Finally, one may prove the quantum coherence for the uncorrelated $N$-particle state ${\rho_N} = {\rho_t^{\otimes N}}$, i.e., the coherence of the entire system, is given by 
\begin{equation}
\label{eq:coherence00001002}
{C}({\rho_t^{\otimes N}}) = {\left[1 + {C}({\rho_t}) \right]^N} - 1 ~.
\end{equation}
In particular, for the case in which the single-particle quantum coherence is much smaller than one, i.e., ${C}({\rho_t}) \ll 1$, the $N$-particle quantum coherence in Eq.~\eqref{eq:coherence00001002} approximately becomes ${C}({\rho_t^{\otimes N}}) \approx N {C}({\rho_t})$.

In principle, the result given by Eq.~\eqref{eq:coherence00002} suggests that, when experimentally measuring the intensity $I(t)$, one could infer the quantum coherence of a single two-level atom of the system. For example, this link would be of particularly interest for experimental setups in which it is unfeasible to address atoms individually. It is important to notice that our results are based on Eq.~\eqref{mecoll} following the mean-field approximation ($N \gg 1$), which refers to a toy model. However, it is straightforward to show that our results are also achieved, still for $N\gg1$, even when we consider individual atomic decay and dephasing as small perturbations in the collective decay. Importantly, it is possible that such constraints do not hold in certain experimental implementations of the referred physical setting, thus entailing some additional complications to obtain a direct relation between the single-atom coherence and the intensity of the emitted radiation. We point out that this deserves further investigation.

\section{Quantum Speed Limit}
\label{sec:QSLsection0001}

In this section we briefly introduce the quantum speed limit (QSL) time and relate it to the superradiance phenomenon. Quantum mechanics imposes a thre\-shold on the minimum evolution time required for a system evolve between two given quantum states, which in turn is certified by the QSL~\cite{1945_JPhysURSS_9_249,1992_PhysicaD_120_188,2009_PhysRevLett_103_160502}. Given a unitary evolution of pure states $|{\psi_0}\rangle$ and $|{\psi_{\tau}}\rangle$ generated by a time-independent Hamiltonian $H$, Mandelstam and Tamm (MT)~\cite{1945_JPhysURSS_9_249} have proved the QSL bound $\tau \geq \hbar\arccos(|\langle{\psi_0}|{\psi_{\tau}}\rangle|)/\Delta E$, in which ${(\Delta E)^2} = \langle{\psi_0}|{H^2}|{\psi_0}\rangle - {\langle{\psi_0}|{H}|{\psi_0}\rangle^2}$ stands for the variance of $H$. Later, Margolous and Le\-vi\-tin (ML)~\cite{1992_PhysicaD_120_188} derived a novel QSL bound for closed quantum systems evolving between two ortho\-go\-nal states, with time-independent Hamiltonian $H$, which reads $\tau \geq \hbar\pi/(2E)$, in which $E = {\langle{\psi_0}|{H}|{\psi_0}\rangle} - {E_0}$ is the mean energy, and $E_0$ the ground state energy of the system. More than a decade after this result, Levitin and Toffoli~\cite{2009_PhysRevLett_103_160502} showed that, by focusing on the case of orthogonal pure states evolving unitarily, the tightest QSL sets ${\tau_{\text{QSL}}} = \max \{\hbar \pi/(2\Delta{E}),\hbar\pi/(2E)\}$. Giovannetti {\it et al.}~\cite{2003_PhysRevA_67_052109} addressed the case of QSLs for mixed states undergoing unitary evolutions, also concluding that entanglement is able to speed up the evolution of composite systems. For more details on QSLs for closed quantum systems, see Refs.~\cite{1983_JPhysAMathGen_16_2993,Aharonov1990,1992_AmJPhys_60_182_Vaidman,1993_PhysRevLett_70_3365,1999_PhysLettA_262_296,2003_36_5587_JPhysAMathGen_Dorje_Brody,2004_PhysicaD_1_189_Luo,2010_PhysRevA_82_022107,2013_JMathPhys_46_335302,Deffner2017,PhysRevA.90.012303_Russell,1506.03199_Mondal_Datta_Sk,PhysRevA.93.052331,Okuyama2018,Modi2018}.

\begin{figure*}[ht]
	\includegraphics[trim = 0mm 0mm 3mm 7mm, clip, width=0.32\linewidth]{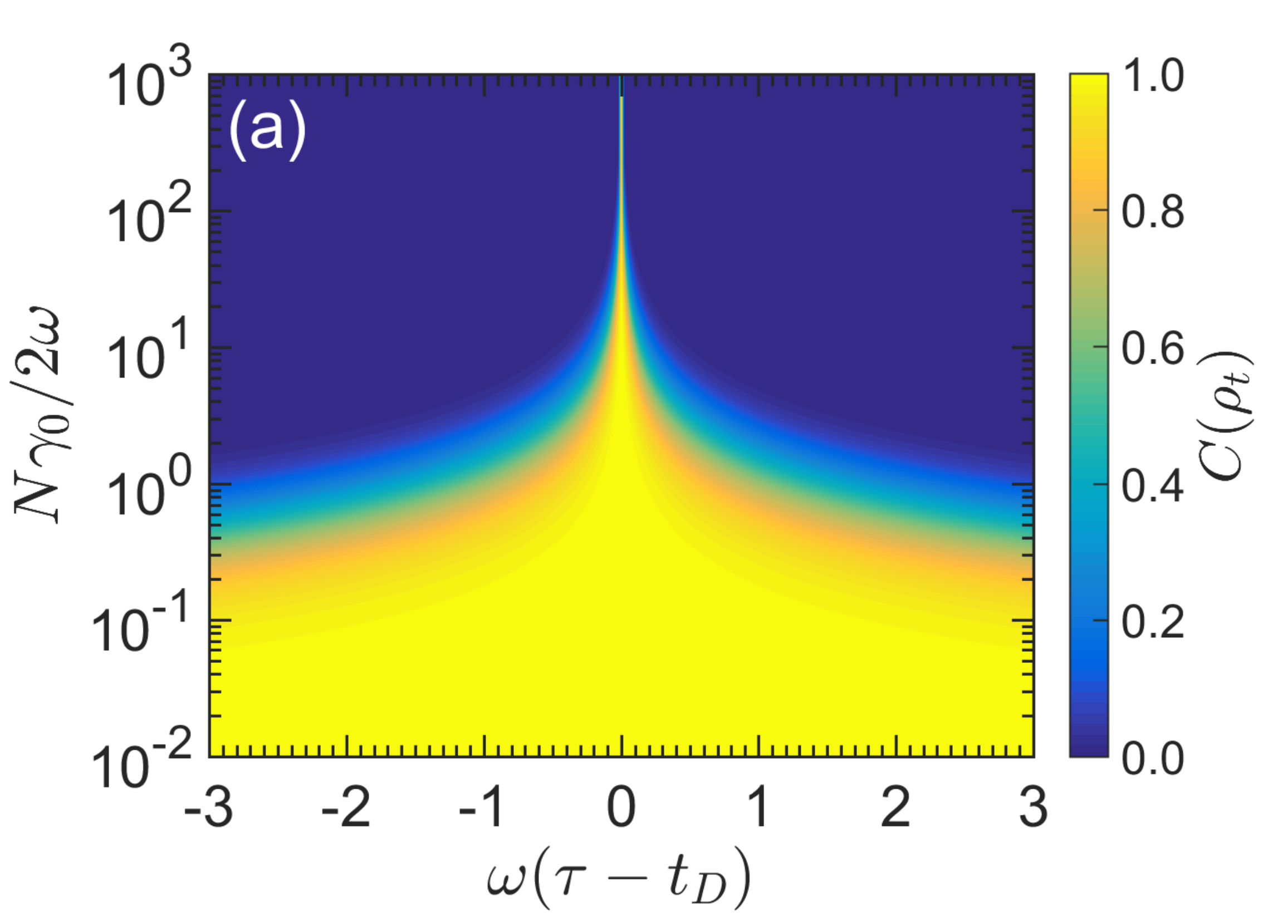}
	\,
	\includegraphics[trim = 0mm 0mm 3mm 7mm, clip, width=0.32\linewidth]{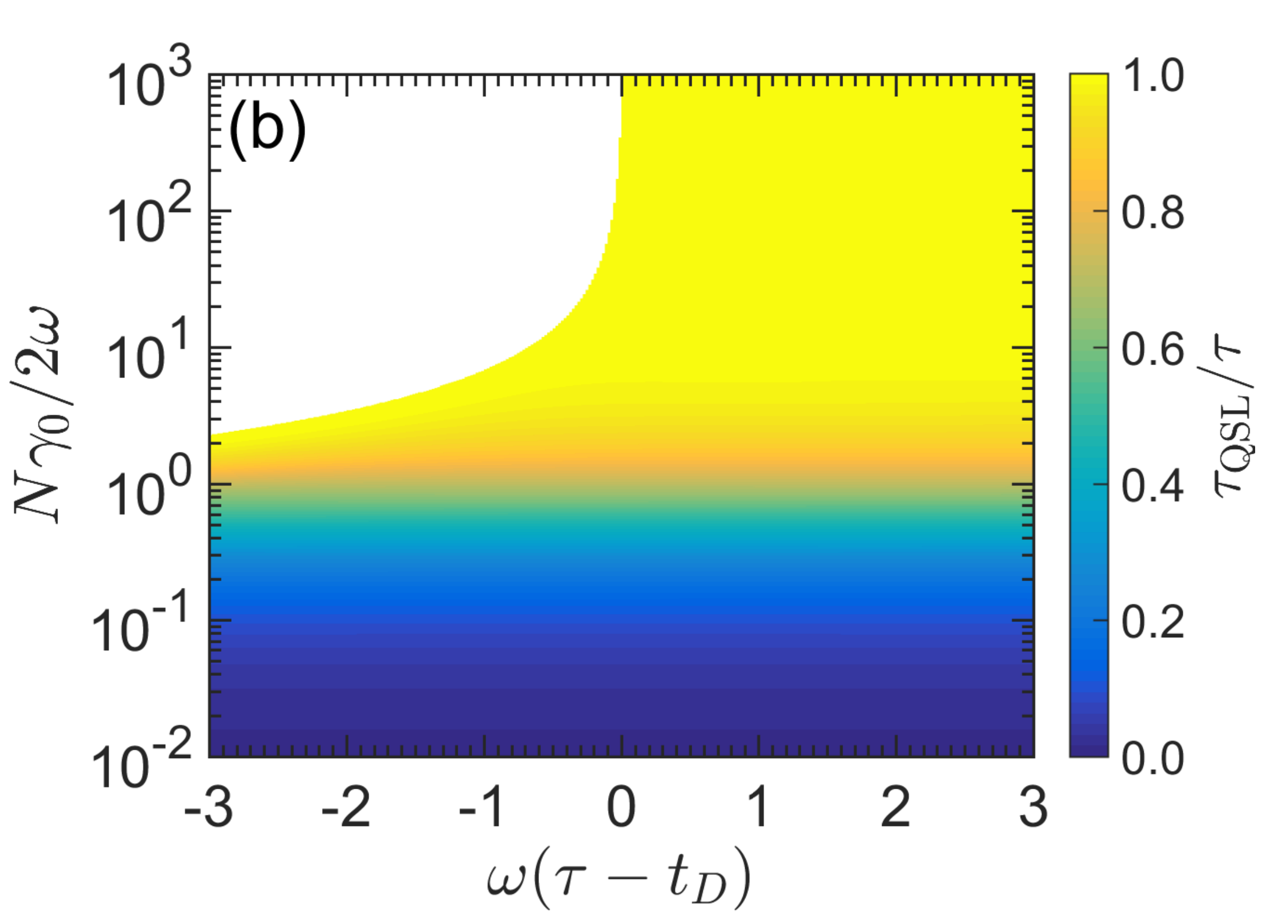}
	\,
	\includegraphics[trim = 0mm 0mm 3mm 7mm, clip, width=0.32\linewidth]{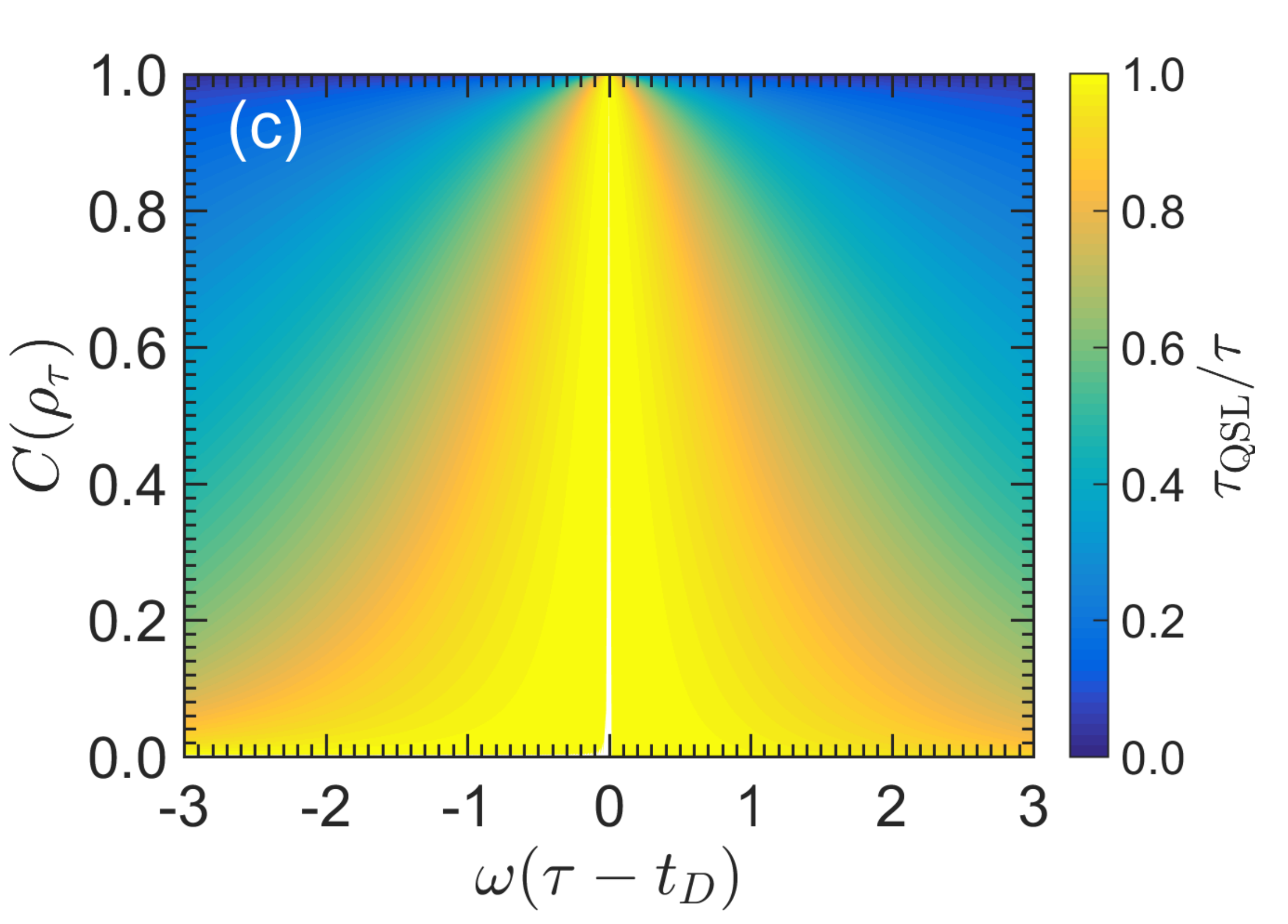}
	\caption{(a) Density plot of quantum coherence ${C}({\rho_t})$ [see Eq.~\eqref{eq:coherence00001}] as a function of the dimensionless parameters $\omega(t - {t_D})$ and the parameter $N{\gamma_0}/(2\omega)$. (b) Density plot of ratio ${\tau_{\text{QSL}}}/\tau$ as a function of $\omega(\tau - {t_D})$ and $N{\gamma_0}/({2\omega})$, for $N = {10^6}$. The white area in (b) is a forbidden region in which the time delay $t_D$ would be negative. (c) QSL ratio ${\tau_{\text{QSL}}}/\tau$ as a function of quantum co\-he\-ren\-ce ${C}({\rho_{\tau}})$ and the parameter $N{\gamma_0}/(2\omega)$ [see Eq.~\eqref{eq:QSLcoherence000xxx001}], fixed the number of atoms $N = {10^6}$.}
	\label{figure}
\end{figure*}

QSL has been also largely investigated for the dynamics of open quantum systems. Indeed, Taddei {\it et al.}~\cite{2013_PhysRevLett_110_050402} and del Campo {\it et al.}~\cite{2013_PhysRevLett_110_050403} have derived the MT bound for arbitrary physical processes, which can be either unitary or nonunitary. Furthermore, Deffner and Lutz~\cite{Deffner2013} derived another class of MT and ML bounds, also showing that non-Markovian signatures can speed up the nonunitary dynamics. Nevertheless, it has been proved the link between speeding up the evolution and non-Markovianity exists only for a certain class of dynamical maps and initial states~\cite{Teittinen_2019}. For completeness, we refer to Refs.~\cite{PhysRevA.89.012307,PhysRevA.91.022102_LiuXuZhu,2015_JPhysAMathTheor_48_045301,PhysRevLett.115.210402,Uzdin_2016,Pires2016,Volkoff2018distinguishability,PhysRevLett.120.070402,PhysRevLett.121.070601,Funo_2019,Garc_a_Pintos_2019,Campaioli2019tightrobust} for other derivations and applications of QSLs for open quantum systems.

In Sec.~\ref{sec:model} we have shown that each single two-level atom of the system undergoes an effective unitary evolution governed by the time-dependent non\-li\-near Hamiltonian $H_t$. The effective two-level system is initialized in the pure state $|{\psi_0}\rangle$, and thus the evolved state $|{\psi_t}\rangle$ will also be pure during the unitary dynamics for any $t \in [0,\tau]$. Therefore, here we will deal with a quantum system undergoing a nonlinear physical process, but still unitary. In this case, the lower bound on $\tau$, which holds for initial and final pure states undergoing a unitary physical process, is obtained from the inequality $\tau \geq {\tau_{\rm QSL}}$, with the QSL time given by~\cite{2013_JMathPhys_46_335302,Deffner2017}
\begin{equation}
\label{tQSL}
{\tau_{\rm QSL}} = \frac{\mathcal{L}(\ket{\psi_0}, \ket{\psi_\tau})}{\overline{\Delta E}_{\tau}} ~,
\end{equation}
where $\mathcal{L}(\ket{\psi_0}, \ket{\psi_\tau}) = \arccos{(|\braket{\psi_0}{\psi_\tau}|)}$ is the Bures angle, i.e., a distance measure between quantum states, while $\overline{\Delta E}_{\tau} = {\tau^{-1}} {\int_{0}^{\tau}} dt \, \Delta {E_t}$ is the time-average of the variance $\Delta {E_t} = \sqrt{\bra{\psi_t}H_t^2 \ket{\psi_t} - \bra{\psi_t}H_t \ket{\psi_t}^2}$ of the time-dependent Hamiltonian $H_t$. In the particular case where $|{\psi_0}\rangle$ is orthogonal to $|{\psi_{\tau}}\rangle$, Eq.~\eqref{tQSL} reduces to ${\tau_{\rm QSL}}  = \pi/(2{\overline{\Delta E}_{\tau}})$.

Physically, ${\tau_{\rm QSL}}$ sets the minimal time the system requires to evolve between states $|{\psi_0}\rangle$ and $|{\psi_{\tau}}\rangle$, also presenting a geometric interpretation discussed as follows. On the one hand, the unitary evolution of $|{\psi_t}\rangle$ des\-cribes an arbitrary path in the manifold of pure states for $t \in [0 ,\tau ]$, thus connecting states $|{\psi_0}\rangle$ and $|{\psi_{\tau}}\rangle$. The length of this path, which generally is not the shor\-test one with respect to the set of paths drawn by $|{\psi_t}\rangle$, is written as ${\int_{0}^{\tau}} dt \, \Delta {E_t}$ and depends on the variance of the Hamiltonian $H_t$, which in turn is nothing but the quantum Fisher information metric for the case of pure states. On the other hand, Bures angle describes the length of the geodesic path connecting states $|{\psi_0}\rangle$ and $|{\psi_{\tau}}\rangle$, and is a function of the overlap of both states. It is quite re\-mar\-ka\-ble that the Bures angle plays the role of a distinguishability measure of quantum states, and stands as the geodesic distance regarding to the quantum Fisher information metric. For a detailed discussion of geometric QSLs, by exploiting the family of Riemannian information metrics defined on the space of quantum states, which in turn encompasses open and closed quantum systems and pure and mixed states, see Ref.~\cite{Pires2016}.

%%%%%%%%
%\begin{figure}[!t]
%\includegraphics[scale=0.375]{FIG02.png}
%\caption{Ratio ${\tau_{\text{QSL}}}/\tau$ as a function of $l_1$-norm of co\-he\-ren\-ce $C({\rho_{\tau}})$, and the difference $\omega(\tau - {t_D})$, for $N = {10^6}$.}
%\label{figure02}
%\end{figure}
%%%%%%%%
Now we will discuss the role played by the superradiance phenomenon and the collective excitations of Dicke states into the QSL time. In order to see this, we will first proceed with the analytical calculation of QSL ratio ${\tau_{\rm QSL}}/\tau$ in Eq.~\eqref{tQSL}. Given that $\bra{\psi_t}H_t^2 \ket{\psi_t} = \left({\omega}/{2}\right)^2 + {\left({N\gamma_0}/{2}\right)^2} {p_t}(1 - {p_t})$, and $\bra{\psi_t}H_t \ket{\psi_t} = ({\omega}/{2})(2{p_t}-1)$, one obtains the time-average of variance as
\begin{widetext}
\begin{equation} 
\label{lhs}
%{\overline{\Delta E}_{\tau}} = \frac{\sqrt{1 + {\alpha^2}}}{2{\alpha}\tau}  \, \left\{ \text{sign}(\tau - {t_D})\arccos{ \left[\text{sech}{\left(\alpha\omega\, (\tau - {t_D})\right)} \right] } + \arccos{ \left[\text{sech}{\left(\alpha\omega\, {t_D}\right)} \right] }\right\} ~,
{\overline{\Delta E}_{\tau}} = \frac{1}{2\tau}\sqrt{1 + {\alpha^{-2}}}  \, \arccos{\left[(1-2p_\tau)(1-2p_0) + 4 \sqrt{p_0 p_\tau (1-p_0)(1-p_\tau)} \, \right]} ~,
\end{equation} 
with $\alpha := N{\gamma_0}/(2\omega)$, while the Bures angle becomes
\begin{equation} 
\label{rhs}
\mathcal{L}(\ket{\psi_0}, \ket{\psi_\tau}) = \frac{1}{2} \arccos{\left[(1-2p_\tau)(1-2p_0) + 4 \sqrt{p_0 p_\tau (1-p_0)(1-p_\tau)} \cos(\omega \tau)\right]  } ~.
\end{equation} 
\end{widetext}
%In particular, we have
%\begin{equation}
%\frac{\tau_{\text{QSL}}}{\tau} =  \frac{2\alpha\, \text{arccos}\left(\sqrt{ \frac{1 +\delta(\tau,{t_D}) - 2\, {C_{l_1}}({\rho_0}){C_{l_1}}({\rho_{\tau}})\, {\sin^2}(\omega\tau/2)  }{2}} \, \right)}{\sqrt{1 + {\alpha^2}}\, \text{arccos}\left(\delta(\tau,{t_D})\right)}
%\end{equation}
%with
%\begin{align}
%\delta(\tau,{t_D}) &:= {C_{l_1}}({\rho_0}){C_{l_1}}({\rho_{\tau}}) \nonumber\\
%&- \text{sgn}(\tau - {t_D})\sqrt{(1 - {{C_{l_1}}({\rho_0})^2})(1 - {{C_{l_1}}({\rho_{\tau}})^2})} ~.
%\end{align}

%%%%%%%%
%\begin{figure}[!thb]
%\includegraphics[scale=0.375]{FIG03.png}
%\caption{Ratio ${\tau_{\text{QSL}}}/\tau$ as a function of $N{\gamma_0}/({2\omega})$ and $\omega(\tau - {t_D})$.}
%\label{figure03}
%\end{figure}
%%%%%%%%

Next, we discuss some remarkable features involving the QSL of the evolution due to superradiant transitions between Dicke states. The state $|{\psi_{\tau}}\rangle$ becomes maximally distinguishable, i.e., orthogonal, to the initial state $|{\psi_0}\rangle$, in the limit $\alpha \to \infty$ (for $\tau > t_D \neq 0$) or $\tau \to \infty$. This means the maximum distinguishability is attainable if the time of evolution exceeds the time delay $t_D$, i.e., when the system populate its superradiant state. Moreover, the bound $\tau \geq {\tau_{\text{QSL}}}$ saturates in the limit $\alpha \to \infty$, as long as $t_D \ne0$ ($p_0 \to 1$). To summarize, QSL time saturates (i) in the so-called overdamped regime, i.e., $N\gamma_0/2\omega \gtrsim 1$; or (ii) if one increases the number of atoms $N$ in the system, maintaining the ratio $\gamma_0/\omega$ fixed. Physically, the system evolves along the geodesic path connecting $|{\psi_0}\rangle$ and $|{\psi_{\tau}}\rangle$ in the mani\-fold of pure quantum states. For instance, this result suggests that quantum information-processing devices based on Dicke superradiance could operate at maximum speed as long as $N \gamma_0/2\omega \gg 1$.

Let us now discuss the role played by quantum coherence into the QSL time. Without loss of generality, here we will set the number of atoms as $N = {10^6}$. It has been proved that entanglement~\cite{2003_PhysRevA_67_052109} can promote a speed up in the time evolution of a quantum system, while the quantum coherence also plays a non-trivial role on the time evolution~\cite{1506.03199_Mondal_Datta_Sk,PhysRevA.93.052331,Pires2016}. Figure~\ref{figure}(b) shows the density plot for the ratio ${\tau_{\text{QSL}}}/\tau$ as a function of dimensionless parameters $\omega(\tau - {t_D})$ and $N{\gamma_0}/(2\omega)$. On the one hand, the inequality $\tau \geq {\tau_{\rm QSL}}$ suddenly saturates in the overdamped regime $N{\gamma_0}/(2\omega) \gtrsim 1$, region in which ${C}({\rho_t}) \to 0$ when $t \ne t_D$, as depicted in Fig.~\ref{figure}(a). On the other hand, in the very underdamped regime $N{\gamma_0}/(2\omega) \ll 1$ where ${C}({\rho_t}) \sim 1$, the ratio ${\tau_{\text{QSL}}}/\tau$ approaches zero, which in turn implies a speed up into the evolution of the two-level system. Therefore, for a fixed difference $\tau - {t_D}$, the ratio ${\tau_{\rm QSL}}/\tau$ decreases as the coherence increases [see Fig.~\ref{figure}(c)], i.e., quantum coherence speeds up the dy\-na\-mics. In other words, the more coherence, the faster the evolution in the mean-field Dicke model of superradiance. Note that the QSL time saturates whenever $C({\rho_{\tau}}) \approx 0$, or close to the peak of the superradiance intensity, which is marked by the delaying time $\tau \approx {t_D}$. It is worth mention that the aforementioned relation between QSL and quantum coherence can be directly observed by rewriting Eqs.~\eqref{lhs} and~\eqref{rhs} in terms of ${C}({\rho_\tau})$, such that

%\begin{widetext}
%	\begin{equation}
%	\label{eq:QSLcoherence000xxx001}
%	\frac{\tau_{\text{QSL}}}{\tau} = \frac{\frac{1}{2}\arccos{\left({C_{l_1}}({\rho_0}){C_{l_1}}({\rho_{\tau}})\cos{(\omega \tau)} - \text{sgn}(\tau - {t_D})\sqrt{(1 - {{C_{l_1}}({\rho_0})^2})(1 - {{C_{l_1}}({\rho_{\tau}})^2})} \, \right)}}{\frac{1}{2}\sqrt{1 + {\alpha^{-2}}}\, \arccos{\left({C_{l_1}}({\rho_0}){C_{l_1}}({\rho_{\tau}}) - \text{sgn}(\tau - {t_D})\sqrt{(1 - {{C_{l_1}}({\rho_0})^2})(1 - {{C_{l_1}}({\rho_{\tau}})^2})} \, \right)}} \,.
%	\end{equation}
%\end{widetext}
\begin{widetext}
	\begin{equation}
	\label{eq:QSLcoherence000xxx001}
	\frac{\tau_{\text{QSL}}}{\tau} = \frac{\frac{1}{2}\arccos{\left({C}({\rho_0}){C}({\rho_{\tau}})\cos{(\omega \tau)} - \text{sgn}(\tau - {t_D})\sqrt{(1 - {{C}({\rho_0})^2})(1 - {{C}({\rho_{\tau}})^2})} \, \right)}}{\frac{1}{2}\sqrt{1 + {\alpha^{-2}}}\, \arccos{\left({C}({\rho_0}){C}({\rho_{\tau}}) - \text{sgn}(\tau - {t_D})\sqrt{(1 - {{C}({\rho_0})^2})(1 - {{C}({\rho_{\tau}})^2})} \, \right)}} \,.
	\end{equation}
\end{widetext}

%%%%%%%%%%%%%%%%%%%%%%%%%%%%%%%%%%%%%%%%%%%%%%%
%%%%%%%%%%%%%%---CONCLUSIONS---%%%%%%%%%%%%%%%%%%%%%%
%%%%%%%%%%%%%%%%%%%%%%%%%%%%%%%%%%%%%%%%%%%%%%%

\section{Conclusions}
\label{sec:conclusions0001}

Dicke superradiance, a phenomenon triggered by the collective coupling of atomic levels with the electromagnetic field, is a subject of wide interest in quantum optics and condensed matter physics, not only from the viewpoint of fundamentals of physics, but also for its promising application in the devising of quantum devices.

In this work we discussed the quantifying of quantum coherence in the Dicke model of superradiance, under the mean-field approximation description, particularly focusing on the $l_1$-norm of coherence. We found that, for all $t \neq {t_D}$, the more particles in the atomic ensemble, the less quantum coherence is stored in the state of the system. It is noteworthy that quantum coherence exhibits its maximum value at time delay $t_D$, for any number $N$ of atoms. Furthermore, we show that quantum coherence is related to the radiation intensity, and thus the former stand as a useful figure of merit to investigate the superradiance phenomenon in the mean-field approach. Due to the aforementioned link, we point out the intensity of radiation emitted by the system vanishes when quantum coherence is identically zero. Remarkably, this result suggests probing the single-atom coherence through the radiation intensity in superradiant systems, which might be useful in experimental setups where is unfeasible to address atoms individually.

%On the one hand, $N$-atom quantum coherence increases exponentially in terms of the respective single-atom coherence measure [see Eq.~\eqref{eq:coherence00001002}]. On the other hand, the multiparticle relative entropy of coherence factorizes itself as the product of relative entropy of coherence of single-atom marginal state. 
%In turn, one may recast time-derivative of relative entropy of coherence in terms of the average radiation intensity emitted by the system of $N$ atoms.

In addition, given the time-dependent single-atom state describing the effective dynamics of each two-level atom [see Eq.~\eqref{hamilteffective0002}], we address the quantum speed limit (QSL) time ${\tau_{\text{QSL}}}$ regarding the unitary evolution between two non-orthogonal pure states. We observed that the in\-crea\-sing of the number of atoms $N$ in the system implies the saturation of the QSL time, thus indicating that Dicke-superradiance-based quantum devices could operate at maximum speed as long as $N \gg 1$. Finally, since QSL time can be recast in terms of the $l_1$-norm of co\-he\-ren\-ce [see Eq.~\eqref{eq:QSLcoherence000xxx001}], we have seen the QSL ratio ${\tau_{\rm QSL}}/\tau$ decreases as the quantum coherence increases, thus concluding that the quantum coherence is a resource that speeds up the o\-verall dy\-na\-mics of the superradiant system.

Our findings unveil the role played by quantum coherence, quantum speed limit, and their interplay in superradiant systems, which in turn could be of interest for the communities of condensed matter physics and quantum optics, particularly for further investigation of Dicke-superradiance-based quantum devices, and for an analysis which goes beyond the mean-field approximation~\cite{PhysRevA.97.052304,PhysRevResearch.2.023125}.

%%%%%%%%%%%%%%%%%%%%%%%%%%%%%%%%%%%%%%%%%%%%%%%
%%%%%%%%%%%%%%---ACKNOWLEDGEMENTS---%%%%%%%%%%%%%%%%%%
%%%%%%%%%%%%%%%%%%%%%%%%%%%%%%%%%%%%%%%%%%%%%%%

\section*{Acknowledgements}

This work was supported in part by the ``EDITAL FAPEPI/MCT/CNPq N$^{\text{o}}$~007/2018: Programa de Infraestrutura para Jovens Pesquisadores/Programa Primeiro Projetos (PPP)''. The authors would like to acknowledge financial support from the Brazilian mi\-nis\-tries MEC and MCTIC, funding agencies CAPES and CNPq, and the Brazilian National Institute of Science and Te\-chno\-logy of Quantum Information (INCT/IQ). This study was financed by the Coordena\c{c}\~{a}o de Aperfei\c{c}oamento de Pessoal de N\'{i}vel Superior -- Brasil (CAPES) -- Finance Code 001.

\end{document}